\documentclass[showpacs,aps,superscriptaddress,prb,twocolumn,amsmath]{revtex4}
\usepackage{amssymb}
\usepackage{graphicx,subfigure,float}
\usepackage[percent]{overpic}
\usepackage{dcolumn}
\usepackage{bm}
\usepackage{color}
\usepackage{hyperref}

\begin{document}

\preprint{APS/123-QED}

\title{Two Dimensional Antiferromagnetic Chern Insulator NiRuCl$_6$}
\author{P. Zhou}
 \affiliation{Key Laboratory of Low-dimensional Materials and Application Technology, School of Material Sciences and Engineering, Xiangtan
University, Xiangtan 411105, China}
\author{L. Z. Sun}
 \email{lzsun@xtu.edu.cn}
 \affiliation{Hunan Provincial Key laboratory of Thin Film Materials
and Devices, School of Material Sciences and Engineering, Xiangtan
University, Xiangtan 411105, China}
\date{\today}
\begin{abstract}
\indent Based on DFT and Berry curvature calculations, we predict that quantum anomalous hall effect (QAHE) can be realized in two dimensional anti-ferromagnetic (AFM) NiRuCl$_6$ with zero net magnetic moment. By tuning spin-orbits coupling (SOC), we find that the topological properties of NiRuCl$_6$ come from its energy band reversal. The results indicate that NiRuCl$_6$ behaves as AFM Chern insulator and its spin-polarized electronic structure and strong spin-orbits coupling (SOC) are the origin of QAHE. Considering the compatibility between AFM and insulator, AFM Chern insulator is more suitable to realize high temperature QAHE because generally N\'{e}el temperature of AFM systems is more easily improved than Curie temperature of ferromagnetic (FM) systems. Due to the different magnetic coupling mechanism between FM and AFM Chern insulator, AFM Chern insulator provides a new way to archive high temperature QAHE in experiments. \\
\end{abstract}
 \maketitle
\indent The quantum anomalous Hall effect (QAHE) arises from the spin-orbit coupling (SOC) showing voltage transverse to the electric current even in the absence of an external magnetic field. Although the fundamental principle of QAHE has been proposed in a honeycomb lattice model\cite{Haldane} long ago, only latest experiment proves that QAHE can be realized in Cr-doped Bi$_2$Se$_3$\cite{xue}. In the system, its ferromagnetic ordering and spin-orbit coupling (SOC) are sufficiently strong to give rise to a topological nontrivial phase with a finite Chern number. However, some extreme experimental conditions including ultra-low temperature hinder its practical application. Searching for materials with QAHE properties working under high temperature is important issue in the scope of condense matter physics. \\
\indent The most direct approach to realize QAHE is to introduce FM order in quantum spin Hall insulators (QSHI) to break its time-reversal symmetry and turn its helical edge states to chiral ones. Such approach requires Chern insulator, or ferromagnetic (FM) insulator with a non-zero Chern number. Unfortunately, finding FM Chern insulator is very challenging due to the fact that there are few ferromagnetic insulators in nature. In conventional diluted magnetic semiconductors, long-range FM order is determined by the Ruderman-Kittel-Kasuya-Yosida(RKKY) interaction which damages QAHE through producing foreign conduction channels. The success observation of QAHE in Cr or V doped (Bi,Sb)$_2$Te$_3$ thanks to the van Vleck mechanism\cite{xue}. According to present experimental situation, besides the nonuniform TM doping, the limitation for improving the Curie temperature is the van Vleck mechanism which is not strong enough to overcome thermal perturbation due to its two order perturbation nature.\\
\begin{figure}
\includegraphics[trim={0.0in 0.8in 0.2in 0.0in},clip,width=3.5in]{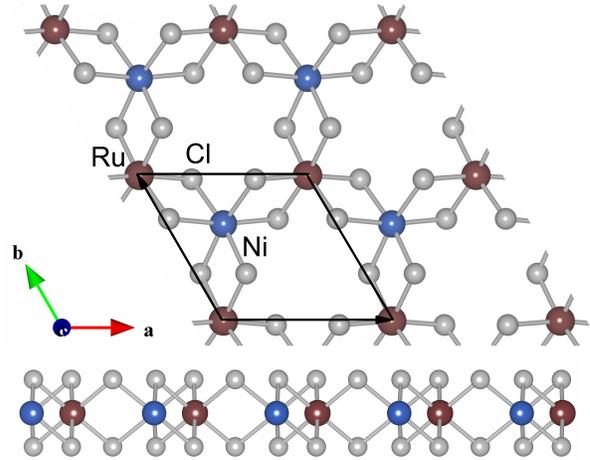}\\
\caption{(color online) Crystal structure for single layer $NiRuCl_6$ from the top (a) and side (b).}\label{fig1}
\end{figure}
\indent As mentioned above, to archive QAHE, researchers pay more attentions to break the time reversal symmetry of topological insulator by importing ferromagnetic order. This reality easily make people produce illusion that QAHE must connect with ferromagnetism. In present letter, taking two dimensional transition metal halides (TMHs) as prototype model, we prove that QAHE can be realized in AFM material as long as the material shows spin-polarization and its time reversal symmetry is broken. The 2D TMHs mainly composed of transition metal atoms and halogen atom with the general formula M$_n$Y$_m$, where M is a transition metal and X is halogen element(Cl, Br and I). To produce spin-polarization, we construct a new type 2D TMHs of M$_1$M$_2$Y$_6$, where the M$_1$ and M$_2$ represent 3\emph{d} and 4\emph{d} TM atom respectively, Y is Cl. Our results indicate that under 5\% compress strain in plane the single layer NiRuCl$_6$ is 2D AFM QAHE insulator, or AFM Chern insulator. \\
\indent Moreover, the pristine single layer NiRuCl$_6$ is a half-metal AFM (HMAFM)\cite{ourarticle1, HMAFM} with zero magnetic moment. Manipulation of spin degrees of freedom in electronic devices to realize information storage, logical calculation, and other functional devices is the premise and main task of spintronics. However, the materials with spin-polarization such as half-metal (HM) and half-semiconductor (HS) often show ferromagnetic order which will restrict the generation of spin-polarized current due to the magnetic domains and stray field accompanied FM order. Moreover, the Curie temperatures of HM and HS are usually below the room temperature due to the generally weak FM super-exchange interaction. Using spin-polarized AFM materials, such as HMAFM, is a feasible solution of such problem. The magnetic ground state of NiRuCl$_6$ is AFM, whereas its VBM and CBM are spin polarized, which is an excellent candidate for spin-polarized AFM materials.\\
\begin{figure}
\includegraphics[width=3.0in]{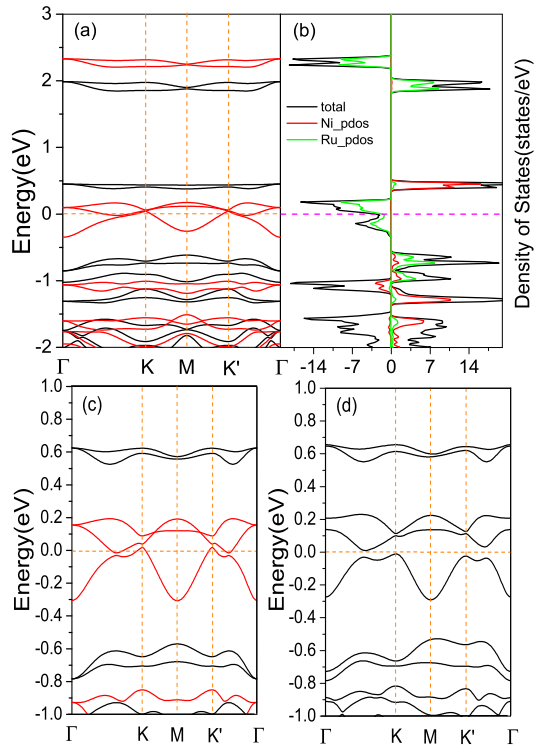}\\
\caption{(color online) Energy band (a) and density of states (b) for pristine $NiRuCl_6$. (c) and (d) are the energy band under 5\% compress stress without and with SOC, respectively.}\label{fig2}
\end{figure}
\begin{figure}
\includegraphics[width=3.5in]{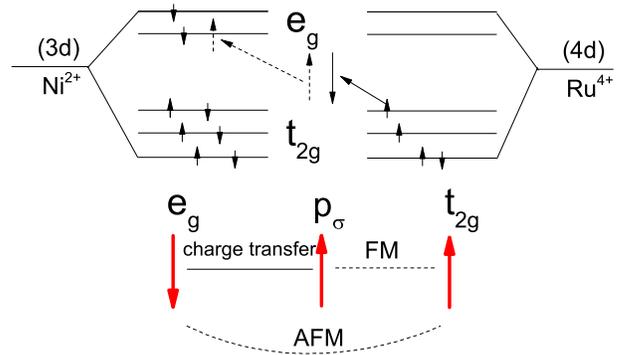}\\
\caption{(color online) Schematic representation of the quasi-superexchange interaction between Ni$^{2+}$ - Cl-3p$_{\sigma}$ - Ru$^{4+}$ in the 90$^\circ$ case. }\label{fig3}
\end{figure}\\
\indent The electronic structures of NiRuCl$_6$ was studied with projector augmented wave (PAW)\cite{PAW} formalism implemented in the Vienna ab initio simulation package\cite{vasp1,vasp2}. The Perdew-Burke Ernzerh of general gradient approximation was used to describe the exchange and correlation functional. The plane-wave cutoff energy was set to be 600 eV and a vacuum space of larger than 15 $\AA$ was set to avoid the interaction between two adjacent layers. The convergence criterion for the total energy was 10$^{-7}$ eV. The crystal lattice and atoms were all relaxed without any restrictions until the Hellmann-Feynman force on each atom was smaller than 0.01 eV/\AA. \\
\indent The structure of  NiRuCl$_6$ as shown in Fig. \ref{fig1} is derived from that of single layer FeCl$_3$\cite{exp1,exp2,exp3} through replacing the two Fe atoms in the primitive cell with Ni and Ru atoms, respectively. Ni and Ru are surround by a distorted octahedron of Cl atoms. We calculate phonon spectrum of the structure and there are no imaginary frequency, which confirms its dynamic stability. The total energy results indicate that NiRuCl$_6$ under AFM state (the spin directions of Ni and Ru are reversal) is 54 meV more stable than that under FM state (the spin directions of Ni and Ru are identical). So the magnetic ground state of NiRuCl$_6$ is AFM and its exchange coupling constant in the Heisenberg model is negative.\\
\begin{figure}
\includegraphics[width=3.5in]{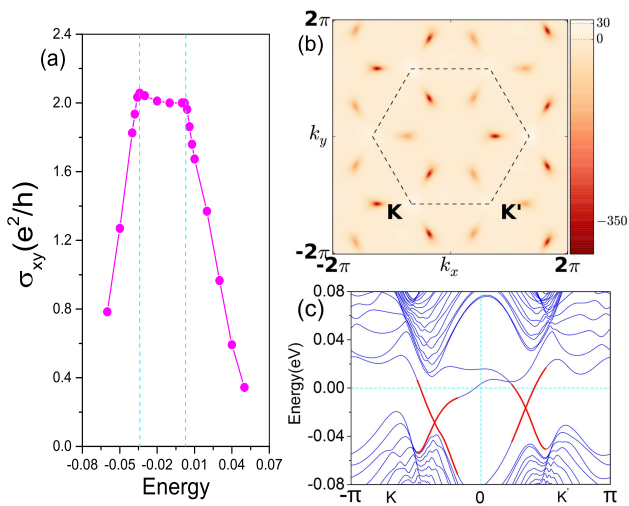}\\
\caption{(color online) (a) Anomalous Hall conductivity when we shift the Fermi level around its original Fermi level; (b) The distribution of the Berry curvature in momentum space for NiRuCl$_6$; (c) Energy band of 20 unit cell zigzag nanoribbon of NiRuCl$_6$. All the results are obtained under 5\% compress strain.}\label{fig4}
\end{figure}
\indent We plot energy band and density of states (DOS) of NiRuCl$_6$ under AFM state in Fig.\ref{fig2}(a). The results show that 2D NiRuCl$_6$ is a kind of typical zero magnetic moment spin-polarized HM, namely the total magnetic moment in the unit cell is zero and the electron around fermi level is 100\% spin-polarized. Through closely inspecting the projected DOS (PDOS) of the system, we find that the three energy bands around Fermi level mainly come from the 4\emph{d} orbitals of Ru. We know that under octahedron surrounding the \emph{d} orbits of TM would be split into $e_g$ and $t_{2g}$. If a further triangle crystal field is present, $t_{2g}$ is decomposed into $a_g$, $e_{g1}^{'}$, and $e_{g2}^{'}$; and $e_g$ breaks into $e_{g1}$ and $e_{g2}$. The PDOS of the system indicates that the three energy bands around Fermi level are $a_g$, $e_{g1}^{'}$, and $e_{g2}^{'}$ of Ru in spin-up channel as shown in Fig.\ref{fig2}, and the Fermi level crosses the energy bands of $e_{g1}^{'}$ and $e_{g2}^{'}$ producing HM nature in the systems. \\
\indent In normal TMCl$_2$ or TMCl$_3$ (TM is 3d transition metal atom) materials, the magnetic order is determined by the superexchange interaction\cite{superex,CrX3}, by which TM atoms interactions follow the GKA rule\cite{GKA1,GKA2} through the intermediate Cl atoms. However, the superexchange builds on highly localized 3\emph{d} TM materials. For NiRuCl$_3$, the magnetic distribution is deviated from normal superexchange due to the existence of more decentralized 4\emph{d} orbits of Ru, we named it quasi-superexchange. We illustrate the quasi-superexchange in Fig.\ref{fig3} where the bond-angle of Ni-Cl-Ru is close to $90^\circ$($96.38^\circ$). Along with the bonding, firstly, 4\emph{s} electrons of Ni transfer to Cl and form Cl$^-$ ion. For the interaction between Cl$^-$ and Ni, the $p_{\sigma}$ of Cl$^-$ is only non-orthogonal to the $d_{x^2-y^2}$ of Ni, forming a partial covalent bond between them. Then the electron hopping will occur from $p_{\sigma}$ orbital of Cl$^-$ to $e_g$ of the Ni$^{2+}$ in the spin-up channel when the system is excited. The remaining spin-down electron of Cl$^-$ with the form of $p_{\pi}$  interacts with the $t_{2g}$ orbital of Ru$^{4+}$.  According to the superexchange mechanism and GKA rule, the magnetic moment of Cl ion should be negative (spin-down), ferromagnetic coupling with Ni$^{2+}$ and anti-ferromagnetic coupling with Ru$^{4+}$. However, our first-principle calculations show that the Cl ion shows tiny positive (spin-up) magnetic moment. The distinct phenomenon can be attributed to the relative delocalization of the 4\emph{d} orbital of Ru in comparison with that of 3\emph{d} of Ni. The conclusion can be verified by the Bader charge, the Bader charge of Cl, Ni, and Ru is 0.365, -0.939, and -1.225e, respectively. Ru lose more electrons than Ni. According to the quasi-superexchange mechanism, the actual spin orientation of Cl atom is decided by the competition between the interaction strength of Cl-Ni and Cl-Ru due to the charge transfer derived from a virtual hopping. In NiRuCl$_3$, the interaction strength between Cl and Ru is larger than that between Cl and Ni which produces $t_{2g}$ orbits of Ru$^{4+}$ transfer more spin-up electrons to Cl atom and finally leads a weak spin-up magnetic moment for Cl. The quasi-superexchange mechanism in  NiRuCl$_3$ produces the anti-ferromagnetic coupling between Ru and Ni in NiRuCl$_3$. The difference of exchange splitting between 3\emph{d} of Ni and 4\emph{d} of Ru produces the 100\% spin-polarization around the Fermi level.\\
\indent Although pristine 2D NiRuCl$_6$ is a typical HMAFM with zero magnetic moment, a global band gap can be opened by applied 3\% to 10\%  compress strain in the plane. The compress strain puts the system transform from zero magnetic moment HMAFM to zero magnetic moment half-semiconductro (HS) AFM. In present work, we chose 5\% compress strain (which can be easily realized in experiment through proper substrate) as example to analyze its topological properties. The band structures of the system under 5\% compress strain in plane are shown in Fig.\ref{fig2} (b) and (c). The results in Fig.\ref{fig2}(b) indicate that under 5\% strain without SOC the system is still HMAFM. When the spin-orbital coupling (SOC) effect is considered and the system is under 5\% strain, it show about 30 meV global band gap, as shown in Fig.\ref{fig2}(c). With the help of our QAHE\cite{ourahe} code and wannier90 code\cite{wannier90}, we calculate the anomalous Hall conductivity with the formula:
\begin{eqnarray}\label{equ1}
\sigma_{xy} = \frac{e^2}{\hbar}\mathcal{C},
\end{eqnarray}
\begin{eqnarray}\label{equ2}
\mathcal{C} = -\sum\limits_n {\int_{BZ}}{\frac{d\bm{k}}{(2\pi)^3}}{f_n}(\bm{k}){\Omega _{n,z}(\bm{k})},
\end{eqnarray}
where $\mathcal{C}$ is the first Chern number, it is finite integer value if QAHE can be realized in the material. $\Omega _{n,z}(\bm{k})$ is the Berry curvature in reciprocal space, we can get the Chern number from the integral of Berry curvature in total first Brillouin zone. Our results show that the Chern number is -2 when the Fermi level locates in the band gap. Meanwhile the system remains HSAFM under the 5\% compress strain in the plane, namely, the system behaves as AFM Chern insulator.  Its anomalous Hall conductivity (AHC) as shown in Fig. \ref{fig4}(a) indicates that the QAHE gap can reach 20 meV when 5\% compress strain is applied. The Berry curvature distribution for the QAHE state is displayed in Fig.\ref{fig4} (b). The distribution shows that the main contribution of Berry curvature comes from the high symmetric line $\Gamma$-K and $\Gamma$-K$^{'}$ close to the two valley K and K$^{'}$ in the reciprocal space. Due to the space inversion symmetry is broken, K and $K^{'}$ are independent of each other and their Berry curvature localization of the maxima is different. However, the integration of the Berry curvature around the K or $K^{'}$ indicates that $C_K =C_{K^{'}}\approx -1$. It is known that nonzero integer Chern number for an 2D material guarantees the existence of quantized edge states when the material is cut into one-dimensional ribbon. To check whether similar quantized edge states exist for NiRuCl$_3$, we calculated the edge states by constructing zigzag nanoribbon based on Maximally Localized Wannier functions Hamilton. The nanoribbon width is 20 unit cell. The results are shown in Fig.\ref{fig4}(c), in which two topologically protected chiral edge states present between the valence band and the conduction band around K and $K^{'}$. The results clearly indicate the Chern number of the system is $C_K =C_{K^{'}} = C/2=-1$.\\
\indent The above results indicate that NiRuCl$_3$ behaves as AFM Chern insulator. To understand the special topological phenomenon, taking K and K$^{'}$ as a reference, we calculate the local orbital projection of wavefunction to label the contribution of $a_g$  and $e_{g1}^{'}$(or $e_{g2}^{'}$) to the three bands around the Fermi level as showed in Fig.\ref{fig5} (The calculation detail can be found in supplementary materials). From low to high energy the three bands are labeled as v1, c1, and c2, respectively. As analysis above, the three 4\emph{d} states of Ru ($a_g$, $e_{g1}^{'}$, and $e_{g2}^{'}$) located around the Fermi level for pristine NiRuCl$_3$ determine its topological properties. The results indicate that around K($K^{'}$) the c2 band mainly comes from $a_g$ state, whereas c1 and v1 bands are nearly equal proportional hybrid by $e_{g1}^{'}$ and $e_{g2}^{'}$ as shown in Fig.\ref{fig5} (a1) and (b1). Because the overlap integration between $a_g$  and $e_{g1}^{'}$(or $e_{g2}^{'}$) under momentum operator must be zero (the details are provided in supplementary materials), the topological properties of NiRuCl$_3$ should be originated from the reversal between $e_{g1}^{'}$ and $e_{g2}^{'}$ due to strong SOC effect. When the SOC is not turned on, large part of the three bands are the equal linear combination of $e_{g1}^{'}$ and $e_{g2}^{'}$ which is depicted as magenta in Fig.\ref{fig5} (a1) and (b1). When the SOC is fully considered, blue ($e_{g1}^{'}$) and red ( $e_{g2}^{'}$) band are clearly separated each other as shown in Fig.\ref{fig5} (a2) and (b2). The separation between  $e_{g1}^{'}$ and $e_{g2}^{'}$ around K(or $K^{'}$) plays a role of energy reversal and leads to poles in Berry curvature space. These poles finally result in an AFM Chern insulator. The $e_{g1}^{'}$ and $e_{g2}^{'}$ own same symmetry and different wavefunction's phase\cite{Ku1}, the band gap must close if they exchange each other. To find out the close of the band gap, we calculate the energy band of NiRuCl$_3$ under different SOC strength and different compress stress in plane as shown in Fig.1s(a). The close of the band gap between $e_{g1}^{'}$ and $e_{g2}^{'}$ is clearly shown in Fig.sa when the SOC strength increased. When SOC is considered, a global energy gap between  $e_{g1}^{'}$ and $e_{g2}^{'}$ can be produced by external compress strain as shown in Fig.1s(a).  The results in Fig.1s(a) indicate that the reversal between $e_{g1}^{'}$ and $e_{g2}^{'}$ and the global energy are remained when the compress strain is within [4\% to 10\%]. When the compress strain is larger than 10\% the above band reversal disappears. \\
\begin{figure}
\includegraphics[width=3.5in]{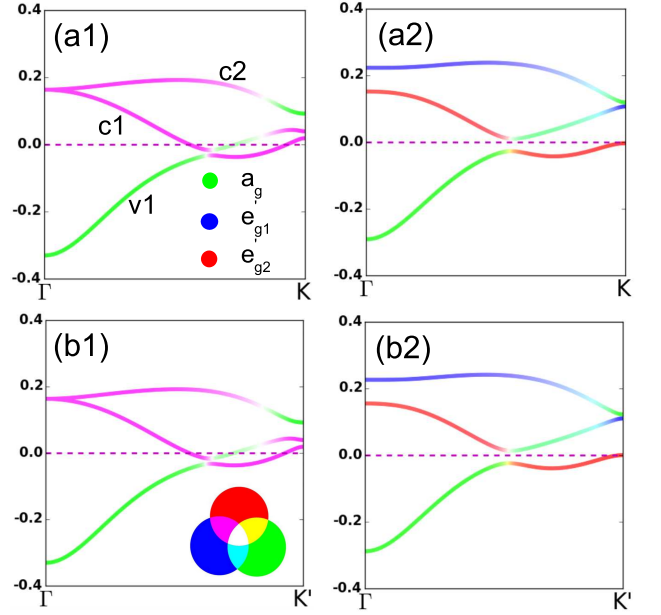}\\
\caption{(color online) The energy band structure along high symmetry line $\Gamma$-K without (a1) and with SOC (a2) and $\Gamma$-$K^{'}$ without(b1) and with SOC(b2). The chromaticity of the three bands derive from the compound through the three primary colours representd $a_g$ (green), $e_{g1}^{'}$ (blue), and $e_{g2}^{'}$ (red). }\label{fig5}
\end{figure}
\indent The N\'{e}el temperature of  NiRuCl$_3$ will limit its future applications in spintronics (HMAFM and HSAFM) and QAHE. To this end, we used monte carlo (MC) and Ising model calculated its N\'{e}el temperature. In the calculations Hamiltonian can be written as $\hat{H}=-\sum_{ij} J  \hat{\textbf{m}}_i \cdot \hat{\textbf{m}}_j$, where $\hat{\textbf{m}}_i$ and $\hat{\textbf{m}}_j$ are the magnetic moments at sites i and j, respectively. The term J is the exchange parameter which is determined by the exchange energy $E_{ex}$ (54meV) with the formula of J = (1/3)$E_{ex}/2m^2$.  Where m =$\mid m\mid$ and the factor of 1/3 is included because there are 3 magnetic coupling interactions in one unit cell. Therefore, J for $NiRuCl_3$ is 2.25 meV. A 50 $\times$ 50 supercell and $10^5$ loop are carried out during the MC simulations. The Specific Heat Capacity(SHC) as the function of temperature T are calculated and the results are showed in Fig.\ref{fig6}. The SHC reaches the highest value when the temperature are close to 120 K suggesting a thermal-induced antiferromagnetism to paramagnetism phase transition. The temperature is several order higher than current experimental temperature. Considering the coupling mechanism of AFM, it is expected show QAHE in experiments under high temperature.\\
\begin{figure}
\includegraphics[width=3.5in]{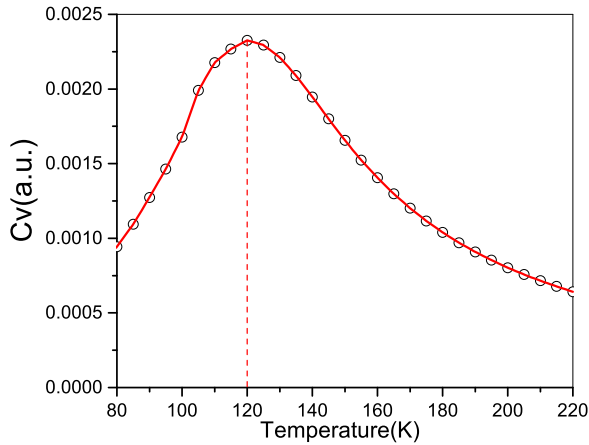}\\
\caption{(color online) The Specific Heat Capacity of NiRuCl$_3$ as the function of temperature T for 50 $\times$ 50 lattices.}\label{fig6}
\end{figure}
\indent  Wehling et al.\cite{Weh} previously pointed out that the electronic structure and magnetic properties of TM doped systems are very sensitive to local Coulomb interactions U. In present letter, we tested the result of U=3.0 and 5.0eV for Ni which does not affect the main conclusion in present work because the energy bands around the Fermi level mainly come from 4\emph{d} orbitals of Ru. We also test the U of Ru up to 2.5 eV, which is close to recent constrained random phase approximation's value of $SrRu_2O6$\cite{Ru2O6}. As the U increases, the maximum of energy band v1 close to the Fermi level shift to $\Gamma$ and concrete analysis is supplied in supplementary materials. Although the different U of Ru lead to different details of electronic structure, NiRuCl$_3$ behaves as AFM Chern insulator.  \\
\indent In summary,  2D NiRuCl$_6$ is an AFM Chern insulator. The quantum anomalous hall effect (QAHE), zero magnetic moment AFMHM, and zero magnetic moment AFMHS can be archived in the system. Its topological properties derive from the reversal of 4d states of Ru around the Fermi level. When external compress strain produce global energy gap, QAHE can be archived in the system. Considering its magnetic coupling is stronger than that of RKKY and van Vleck mechanism, AFM Chern insulator is hoping to realize high temperature QAHE,  AFMHM, and  AFMHS in experiments.  Our results broaden the outlook of Chern insulator and QAHE as well as  pave a new way to realize high temperature QAHE.\\
\begin{acknowledgments}
This work is supported by the National Natural Science Foundation of China (Grant No. 11574260) and scientific research innovation project of Hunan Province(CX2015B219).
\end{acknowledgments}


\begin{thebibliography}{1}

\bibitem{exp1}Bointon, T. H. et al. , Nano Lett. 14, 1751每1755 (2014).
\bibitem{exp2}  Zhan, D. et al. , Adv. Func. Matter. 20, 3504每3509 (2010).
\bibitem{exp3}  Zhao, W., Tan, P. H., Liu, J. and Ferrari, A. C., J. Am. Chem. Soc. 133, 5941每5946 (2011).
\bibitem{Haldane} F. D. M. Haldane, Phys. Rev. Lett. 61, 2015 (1988)
\bibitem{xue}Chang C Z, Zhang J, Feng X, et al., Science 340, 167 (2013).
\bibitem{PAW} Blochl, P. E. Phys. Rev. B 1994, 50, 17953-17979.
\bibitem{vasp1} Kresse, G.; Hafner, J. Phys. Rev. B 1993, 47, 558-561.
\bibitem{vasp2} Kresse, G.; , J. Phys. Rev. B 1996, 54, 11169-11186.
\bibitem{wien2k}P. Blaha, K. Schwarz, G. K. H. Madsen, D. Kvasnicka, and J. Luitz, WIEN2K: An Augmented Plane Wave plus Local Orbitals Program for Calculating Crystal Properties (Vienna University of Technology, Austria, 2001).
\bibitem{FuKane} L.Fu and C.L.Kane;Phys.Rev.B 76,045302(2007)
\bibitem{HMAFM} Xiao Hu, Adv. Mater. 24, 294每29 (2012)
\bibitem{Soluyanov} A.A.Soluyanov and D.Vanderbilt, Phys.Rev.B 83,035108 (2011);Phys.Rev.B 83,235401 (2011).
\bibitem{phonopy} Atsushi Togo, Fumiyasu Oba, and Isao Tanaka, Phys. Rev. B, 78, 134106 (2008).
\bibitem{bader} G. Henkelman, A. Arnaldsson, and H. J\'{o}nsson, Comput. Mater. Sci. 36, 254-360 (2006).
\bibitem{fermi1} Young-Jun Yu, Yue Zhao, Sunmin Ryu, Louis E. Brus, Kwang S. Kim, and Philip Kim, Nano. Lett. 9, 3430 (2009).
\bibitem{fermi2} Chih-Pin Lu, Guohong Li, K. Watanabe, T. Taniguchi, and EvaY. Andrei, Phys. Rev. Lett. 113, 156804 (2014).
\bibitem{fermi3} He Tian, Zhen Tan, Can Wu, Xiaomu Wang, Mohammad Ali Mohammad, Dan Xie, Yi Yang, Jing Wang, Lain-Jong Li, Jun Xu and Tian-Ling Ren, Scientific Reports, 4, 5951 (2014).
\bibitem{ourahe} P. Zhou, L. Z. Sun, arXiv:1507.03457.
\bibitem{ourarticle1} Junjie He, Pan Zhou, N. Jiao, Xiaoshuang Chen, Wei Lu, L. Z. Sun, RSC Adv., 5, 46640-46647(2015).
\bibitem{superex}Suzuki M, Suzuki IS. [October 7, 2011];Lecture Notes on Solid State Physics (superechange interaction) www2.binghamton.edu/physics/docs/super-exchange.pdf.
\bibitem{CrX3} J. Liu, Q. Sun, Y.Kawazoe and P. Jena, Phys. Chem. Chem. Phys., 2015, DOI: 10.1039/C5CP04835D.
\bibitem{GKA1} P.W. Anderson, Chapter 2 (p.25-83), in Magnetism, edited by G.T. Rado and H. Suhl (Acaddemic Press, New York, 1963).
\bibitem{GKA2} J.B. Goodenough, Magnetism and the Chemical Bond (Interscience Publisher, New York, 1963).
\bibitem{wannier90} Wannier90: A Tool for Obtaining Maximally-Localised Wannier Functions A. A. Mostofi, J. R. Yates, Y.-S. Lee, I. Souza, D. Vanderbilt and N. Marzari, Comput. Phys. Commun. 178, 685 (2008).
\bibitem{Weh}T. O. Wehling, A. I. Lichtenstein, M. I. Katsnelson,  Phys. Rev. B, \textbf{84}, 235110(2011).
\bibitem{Ru2O6} W. Tian, C. Svoboda, M. Ochi, M. Matsuda, H. B. Cao, J.-G. Cheng, B. C. Sales, D. G. Mandrus, R. Arita, N. Trivedi, and J.-Q. Yan, Phys. Rev. B, 92, 100404(R) (2015).
\bibitem{Ku1} Tom Berlijn, Dmitri Volja, and Wei Ku, Phys. Rev. Lett. 106, 077005 (2011)
\end{thebibliography}

\end{document}